# Investigation of surface homogeneity of (3200) Phaethon

H.-J. Lee[a,b], M.-J. Kim[b1], D.-H. Kim[a,b], H.-K. Moon[b], Y.-J. Choi[b,c], C.-H. Kim[a], B.-C. Lee[b], F. Yoshida[d], D.-G. Roh[b] and H. Seo[b,e]

[a]*Chungbuk National University, 1 Chungdae-ro, Seowon-Gu, Cheongju, Chungbuk, 28644, Korea*

[b]*Korea Astronomy and Space Science Institute, 776 Daedeukdae-ro, Yuseong-gu, Daejeon, 34055, Korea*

[c]*University of Science and Technology, 217, Gajeong-ro, Yuseong-gu, Daejeon 34113, Korea*

[d]*Planetary Exploration Research Center, Chiba Institute of Technology, 2-17-1 Tsudanuma, Narashino, Chiba, 275-0016, Japan*

[e]*Intelligence in Space, 96 Gajeongbuk-ro, Yuseong-gu, Daejeon, 34111, Korea*

**ABSTRACT**

Time-series multi-band photometry and spectrometry were performed in Nov.-Dec. 2017 to investigate the homogeneity of the surface of asteroid (3200) Phaethon. We found that Phaethon is a B-type asteroid, in agreement with previous studies, and that it shows no evidence for rotational color variation. The sub-solar latitude during our observation period was approximately 55 °S, which corresponded to the southern hemisphere of Phaethon. Thus, we found that the southern hemisphere of Phaethon has a homogeneous surface. We compared our spectra with existing spectral data to examine the latitudinal surface properties of Phaethon. The result showed that it doesn't have a latitudinal color variation. To explain this observation, we investigated the solar-radiation heating effect on Phaethon, and the result suggested that Phaethon underwent a uniform thermal metamorphism regardless of latitude, which was consistent with our observations. Based on this result, we discuss the homogeneity of the surface of Phaethon.

---

[1]  Corresponding author: skarma@kasi.re.kr

# 1. Introduction

(3200) Phaethon (1983 TB) (hereafter "Phaethon") was first discovered by the Infrared Astronomy Satellite (IRAS) in October 1983 (Green & Kowal 1983). Classified as an Apollo-type Near-Earth Asteroid (NEA), Phaethon is an object that approaches very close to the Sun with a perihelion distance of 0.14 AU. In particular, Phaethon is regarded as a parent body of the Geminids meteor stream (Whipple 1983; Fox et al. 1984; Green et al. 1985; Gustafson 1989; Williams & Wu 1993). The parent body of a meteor shower emits dust and a part of the dust reaches to the Earth. Therefore Phaethon is a body providing extraterrestrial materials to the Earth. The activity of Phaethon is very unique and appears to be sporadic. Its cometary activity has not been observed in previous observation data (Urakawa et al. 2002; Hsieh & Jewitt 2005; Kraemer et al. 2005; Wiegert et al. 2008), and has been only detected by the STEREO (Solar Terrestrial Relations Observatory) spacecraft data (Jewitt & Li 2010; Li & Jewitt 2013). Therefore, Phaethon is a rather unique object that is considered to be a comet-asteroid transition object (Licandro et al. 2007). Due to these properties, Phaethon was selected as the target of the DESTINY$^+$ mission of JAXA/ISAS.

Phaethon is a B-type asteroid (Green et al. 1985; Binzel et al. 2001, 2004; Bus & Binzel 2002; DeMeo et al. 2009) and assumed to be a fragment of the large main-belt asteroid, (2) Pallas (de León et al. 2010a). F/B-type asteroids are commonly linked to dehydrated CI/CM chondrites (Hiroi et al. 1993; Hiroi et al. 1996). The spectral features of CI/CM chondrites are chemically altered from hydrated C-type to F/B-type by thermal metamorphism (Tholen 1984; Bell et al. 1989; Hiroi et al. 1993; Hiroi et al. 1996). In other words, a B-type asteroids are considered to have undergone surface alteration by thermal metamorphism. However, we note that other surface properties, such as the average grain size of the surface regolith (Binzel et al. 2015, Vernazza et al. 2016), may also make carbonaceous asteroids to have negative spectral slopes. Ohtsuka et al. (2009) assumed that the heating source that caused the thermal metamorphism on Phaethon's surface is solar-radiation and they investigated its effects. Ohtsuka et al. (2009) calculated its surface temperature caused by solar radiation using the pole solution of Krygly et al. (2002) and the two extreme thermal models: the Near-Earth Asteroid Thermal Model (NEATM; Harris 1998) and the Fast Rotation Model (FRM; Lebofsky & Spencer 1989). They revealed that Phaethon can have a sufficient temperature to dehydrate and decompose surface materials due to solar radiation and that the solar-radiation heating effect is latitude-dependent. They estimated that the northern hemisphere, in particular the arctic region north of 45 °N, must have undergone much more dehydration and decomposition than the southern hemisphere. Furthermore, they attempted to verify their hypothesis by collecting previously observed spectra of Phaethon, but could not confirm their hypothesis with existing dataset.

As mentioned above, Phaethon is an object that is assumed to have undergone changes on its surface by collisional/thermal evolutions since its formation. In particular, since Phaethon's parent body is assumed to have split into Phaethon and (155140) 2005 UD (hereafter 2005 UD) that have very similar orbit with Phaethon's one, approximately ~100 kyr ago (Hanus et al. 2016), it is presumed that some traces from this event are left on the surface. In fact, rotational surface color variations have been observed on 2005 UD (Kinoshita et al. 2007). Since Phaethon and 2005 UD have similar orbits, they must have undergone similar space weathering. Therefore, if the color variation of 2005 UD was caused by fragmentation with Phaethon and subsequent exposure of its fresh surface, it is expected that rotational color variation should also be detected on Phaethon.

To test the above hypothesis, we performed multi-band photometry and visible-spectrometry in Nov. ~ Dec. 2017 to verify any change in its surface properties. In Section 2, we describe the observation and data reduction methods. In Section 3, we discuss the surface properties of Phaethon based on our observations. Section 3.1 discusses the taxonomy type of Phaethon, and Section 3.2 discusses the surface homogeneity. Finally, Section 4 presents a summary and conclusions.

## 2. Observations and data reduction
### 2.1 Observations

In order to examine the surface homogeneity of Phaethon, we conducted a time-series multi-band photometric observations at Mt. Lemon Optical Astronomy Observatory (LOAO) in Arizona, USA from November 11 to 13, 2017. We used the 1m telescope and the 4K x 4K e2v CCD. The pixel scale of the CCD camera was 0.8 arcsec/pix, and the field of view was 28.1 arcmin x 28.1 arcmin. For investigating the suspected color variation according to the rotation of Phaethon, we used Johnson-Cousins BVRI-filters and the exposure time for each image was 100 s. The exposure time was determined to maintain the length of the trailed signal of trailed image of the target within 2" on average, considering its sky motion and apparent magnitude. The predicted seeing was smaller than the seeing in the actual observation results. Therefore, this prediction is reasonable and the asteroid in our observation were observed to be a point source.

At the same time, we carried out a visible-spectrometry to examine the changes in the spectra of Phaethon in more detail. The observation was performed at Mt. Bohyunsan Optical Astronomy Observatory (BOAO) in Korea using the 1.8m telescope, with the 4K x 4K e2v CCD and long-slit spectrograph on December 7, 2017. The width of the slit used here was 300 $\mu m$, and the resolution was 2.4 Å/pix. The observation wavelength was 4,000 ~ 7,000 Å, and the exposure time was 300 s. The detail of geometries and observational circumstances are shown in Table 1.

**Table 1.** The geometries and observational circumstances. *

| Time (UTC) | Δ [AU] | R [AU] | α [°] | V [Mag.] | Sky Motion [″/min] | SITE | Seeing [″] | Exp. Time [s] | Sky condition | Observation type | Tracking Mode | Solar analog |
|---|---|---|---|---|---|---|---|---|---|---|---|---|
| 2017-11-11.4 | 0.695 | 1.496 | 33.1 | 16.1 | 0.23 | LOAO | 3.2 | 100 | Cirrus | Multi-band photometry (BVRI) | Sidereal | - |
| 2017-11-12.4 | 0.675 | 1.485 | 32.9 | 16 | 0.23 | LOAO | 2.7 | 100 | Cirrus | Multi-band photometry (BVRI) | Sidereal | - |
| 2017-11-13.4 | 0.654 | 1.473 | 32.8 | 15.9 | 0.24 | LOAO | 2.7 | 100 | Cirrus | Multi-band photometry (BVRI) | Sidereal | - |
| 2017-12-07.7 | 0.186 | 1.157 | 20.7 | 12.3 | 4.57 | BOAO | 2.9 | 300 | Cirrus | Long-slit spectroscopy | Non-sidereal | HD 28099, HD 25680 and HD 29461 |

*Δ: geocentric distance. *R*: heliocentric distance. *α*: phase angle. V : apparent predicted magnitudes. Exp. Time : exposure time

**2.2 Data reduction**

**2.2.1 Multi-band photometry**

The IRAF/CCDRED package was used to pre-process the observed data. We corrected for BIAS, DARK, and FLAT fielding during this process. The WCS solution of each image was determined by matching with the USNO B1.0 catalog using SCAMP (Bertin 2006). Aperture photometry was performed using the IRAF/APPHOT package. The aperture radius was set equal to the Full Width Half Maximum (FWHM) of the stellar profile in order to have the maximum S/N ratio (Howell 1989). And the inner and outer sky annulus were set to five times and six times the FWHM, respectively, considering the stellar density in the images. We carried out standardization to obtain the color indices of Phaethon. The standardization was executed using the ensemble normalization technique (Gilliland et al. 1988; Kim et al. 1999) with the Pan-STAARS Data Release 1 catalog (PS DR1; Chambers et al. 2016). The magnitude of the SDSS filters of PS DR1 was convoluted to the Johnson-Cousins filter magnitude using the transformation equations proposed by Tonry et al. (2012). The catalog stars between 11 and 16 magnitude in BVRI were used for photometric calibration. The calibrated magnitude error, which considers the zero magnitude error and the instrumental magnitude error, has a value of 0.02-0.05 mag. In addition, the light curve amplitude of Phaethon is less than 0.15 magnitude along its rotation (Hanus et al. 2016; Kim et al. 2018). Because it was rotated ~11-degrees between the time period when our observations were made with the first and the last filter. In this case, the light variation of Phaethon can be assumed to be less than 0.01 magnitude during this period, when presuming a gradual change in its brightness. This is smaller than the photometric error. Hence, the brightness variability was not be considered when calculating the color indices.

**2.2.2 Visible spectrometry**

For pre-processing of the observed spectroscopic data, BIAS, DARK, and FLAT fielding were corrected using the same IRAF/CCDRED package as for photometry. For FLAT fielding, the observation images of the THL lamp were used. In addition, the IRAF/SPECRED package was used for spectrum extraction, wavelength calibration, flux calibration and Doppler shift correction, and a Fe-Ne lamp was used as the wavelength comparison light source. To remove the solar spectrum from the observed spectrum of the asteroid, we additionally observed solar-analog stars HD 28099, HD 25680 and HD 29461 (Hardorp 1978; Hoffleit & Jaschek 1991). We used the solar-analog stars observed at the same air mass as the asteroid to remove solar colors from the asteroid spectrum.

## 3. Results and discussions

### 3.1 Taxonomy type

The existing taxonomic type of Phaethon was verified using the color index obtained through standardization of the multi-band photometry of Phaethon and the spectrum obtained through long-slit spectroscopy. First, we determined the taxonomic type using the color index, for which we used the mean value of the observed color indices. The color indices of Phaethon are shown in Table 2. Based on these indices, we estimated that the taxonomy of Phaethon is B-type using the classification method suggested by Dandy et al. (2003) as shown in Figs. 1 and 2. The taxonomy of Phaethon was additionally confirmed by the acquired spectra in Fig. 3 which show to be typical for any B-type asteroid with a blue slope. It was a previously known (Green et al. 1985; Binzel et al. 2001, 2004; Bus & Binzel 2002; DeMeo et al. 2009) and our new observations are consistent with the existing spectra.

**Table 2.** Color indices of Phaethon

| B – V | V- R | R – I | Ref. |
| --- | --- | --- | --- |
| - | 0.34 | - | Skiff et al. (1996) |
| 0.59 ± 0.01 | 0.35 ± 0.01 | 0.32 ± 0.01 | Dundon (2005) |
| 0.61 ± 0.01 | 0.34 ± 0.03 | 0.27 ± 0.04 | Kasuga et al. (2008) |
| 0.64 ± 0.02 | 0.34 ± 0.02 | 0.31 ± 0.03 | This work |

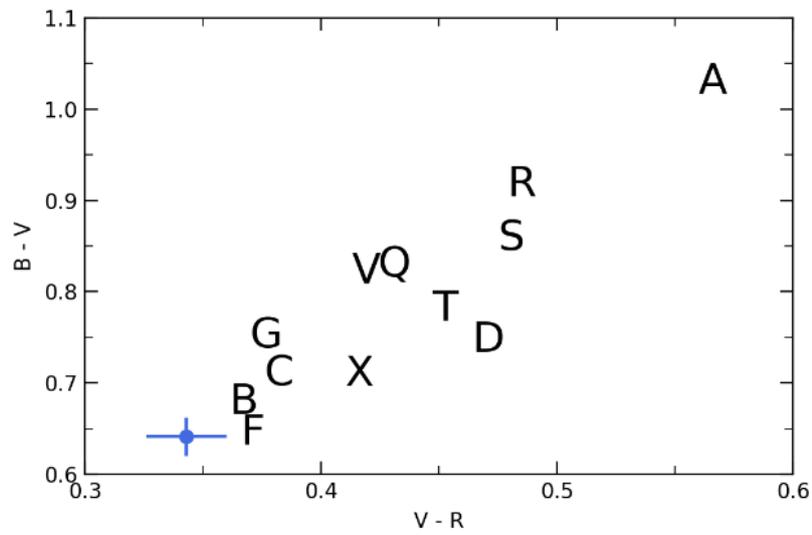

**Figure 1.** V-R vs B-V two-color diagram. The characters indicate the taxonomic types classified by Dandy et al. (2003) and the blue dot indicates Phaethon from our observation.

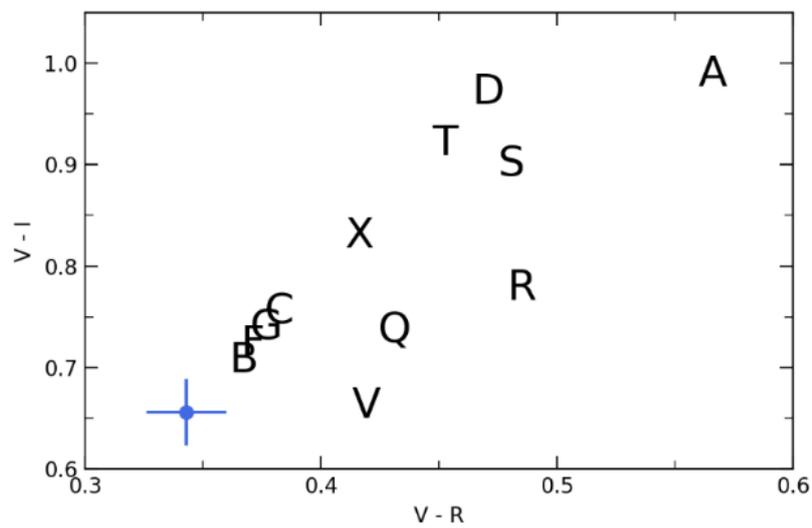

**Figure 2.** V-R vs V-I two-color diagram. The characters and dot have the same meanings as in Fig. 1.

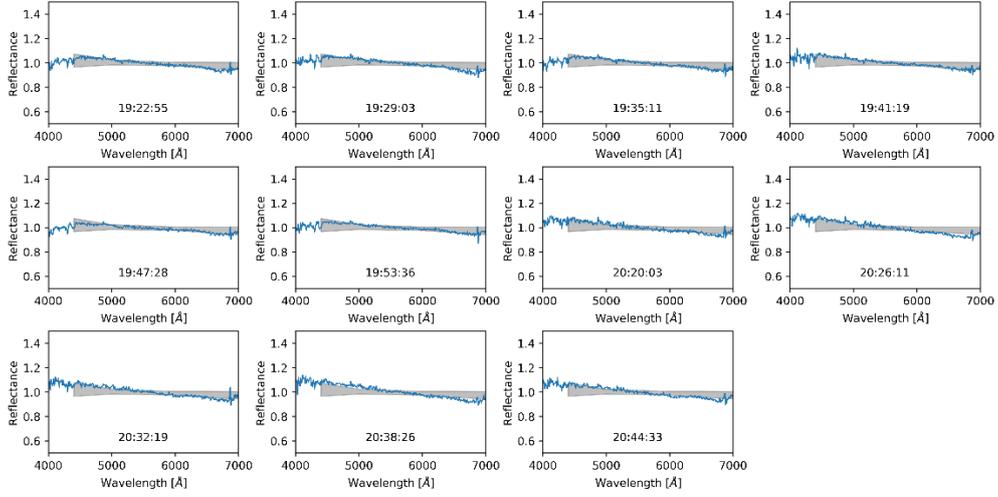

**Figure 3.** The spectra of Phaeton taken on Dec. 07 2017. The gray area corresponds to the range of B-type asteroid spectra (Bus & Binzel, 2002).

**3.2 Surface homogeneity of Phaethon**

Our photometry and spectroscopy were also used in the investigation of the surface homogeneity by converting the photometric data to reflectivity using the color index of the Sun proposed by Ramirez et al. (2012). We used the normalized spectral gradient of reflectivity (S′; Luu & Jewitt 1990) to check whether there are any change in reflectivity. S′ can be determined through the following equation:

$$S'(\lambda_1, \lambda_2) = \frac{dS/d\lambda}{S_{\lambda_c}} \qquad (1)$$

where $dS/d\lambda$ is the change rate of the reflectivity in the range $\lambda_1 < \lambda < \lambda_2$, and $S_{\lambda_c}$ is the reflectivity at $\lambda_c$. $\lambda_c$ denotes a wavelength between $\lambda_1$ and $\lambda_2$. To compare our BVRI photometry with our spectra, we used 4,420 Å, 5,400 Å and 6,470 Å, which are the effective wavelengths of the B, V and R filters, for $\lambda_1$, $\lambda_2$ and $\lambda_c$, respectively. B-type asteroids don't show variation in the spectral gradient as function of phase angle (Lantz et al., 2018). Hence, we didn't correct the phase reddening effect. Then, the apparent geometry of Phaethon at each observation date was determined and defined as sub-earth points using the pole solution and shape model obtained by Kim et al. (2018).

### 3.2.1 Longitude-dependent surface homogeneity

We examined the longitudinal color variation using our photometry and spectroscopy during the observation period. The sub-earth latitude during the observation period was ~55°S; thus, our measured colors correspond to the surface of Phaethon's southern hemisphere. Fig. 4 shows the spectral gradient for the longitude, obtained from the observed data. Fig. 4 reveals that there is no longitude-dependent color variation. This suggests that the southern hemisphere of Phaethon has homogeneous surface properties.

As mentioned in Section 1, if the color variation detected on 2005 UD was caused by fragmentation with Phaethon, a color variation should appear on Phaethon's surface as well. However, no color variation was detected in the observation data for the southern hemisphere. Hence, time-series multi-band photometry and spectra observations are also required during the apparition in which Phaethon's northern hemisphere can be observed.

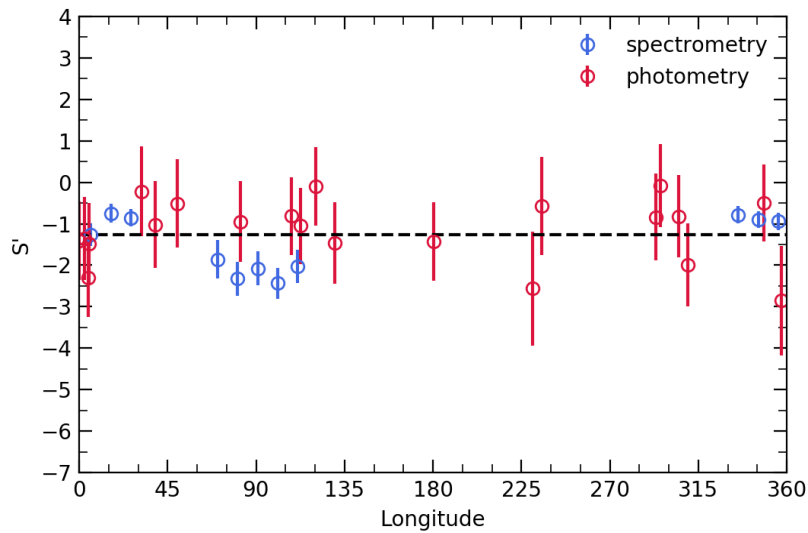

**Figure 4.** Spectral gradient $S'$ according to the longitude of Phaeton. The dots indicate $S'$ determined on the observation data, and the dashed lines represent the mean values of $S'$. The errors in the slope $S'$ for spectroscopy were estimated considering the noise in the spectra.

### 3.2.1 Latitude-dependent surface homogeneity

We collected existing visible spectral data that include un-published data observed by IRTF/SpeX (Rayner et al., 2003) acquired during various apparitions to examine the latitudinal surface homogeneity of Phaethon. As for our observation data, we determined $S'$ from the collected spectra and examined the latitude-dependent color variations. The collected spectra, our own spectral dataset and the geometry at the time of observation are listed in Table 3. In addition, Fig 5. displays Phaethon's geometrical aspects at the time of spectral observation. Fig. 6 does not show significant latitude-dependent color change in the southern hemisphere. At the same time, it appears that slope $S'$ may drop to below average if more of the northern hemisphere is visible. Hence, we may conjecture that the northern hemisphere might have experienced more thermal metamorphism than the southern hemisphere. However, we should be cautious because the deviation is within the error bar.

**Table 3.** The geometries and spectral gradient $S'$ from archival and our spectral datasets.*

| Time (UTC) | $\Delta$ [AU] | $R$ [AU] | $\alpha$ [°] | $\phi_{se}$ [°] | $S'$ [%/$10^3$Å] | Ref. |
|---|---|---|---|---|---|---|
| 1994-11-15 | 0.810 | 1.710 | 20.0 | -35.2 | -1.63 ± 0.15 | Binzel et al. (2004) |
| 2004-12-10 | 0.638 | 1.574 | 17.5 | -8.8 | -1.70 ± 0.15 | SMASS web |
| 2004-12-20 | 0.614 | 1.469 | 29.6 | 0.3 | -1.73 ± 0.10 | de León et al. (2010b) |
| 2014-11-28 | 0.841 | 1.807 | 9.4 | -14.2 | -1.64 ± 0.15 | SMASS web |
| 2014-11-29 | 0.834 | 1.798 | 9.5 | -14.2 | -1.64 ± 0.15 | SMASS web |
| 2014-12-01 | 0.819 | 1.782 | 10.2 | -9.6 | -1.64 ± 0.15 | SMASS web |
| 2017-10-16 | 1.235 | 1.754 | 33.6 | -53.0 | -1.64 ± 0.15 | SMASS web |
| 2017-12-07 | 0.185 | 1.167 | 20.8 | -55.1 | -1.63 ± 0.10 | This work |

*Δ: geocentric distance. $R$: heliocentric distance. $\alpha$: phase angle. $\phi_{se}$: the sub-earth latitude. SMASS web : http://smass.mit.edu/smass.html

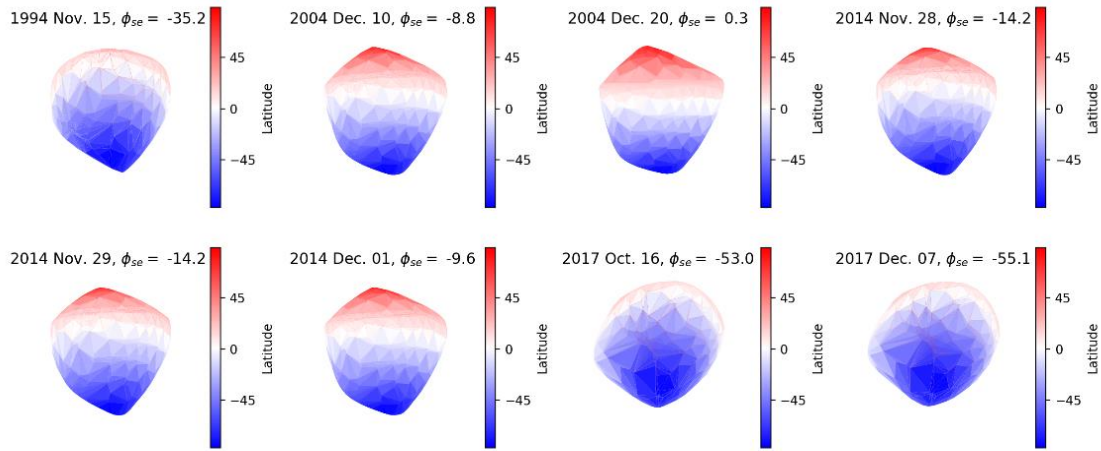

**Figure 5.** The aspects of Phaethon at the time of spectral observation. The Symbol $\phi_{se}$ is the sub-earth latitude.

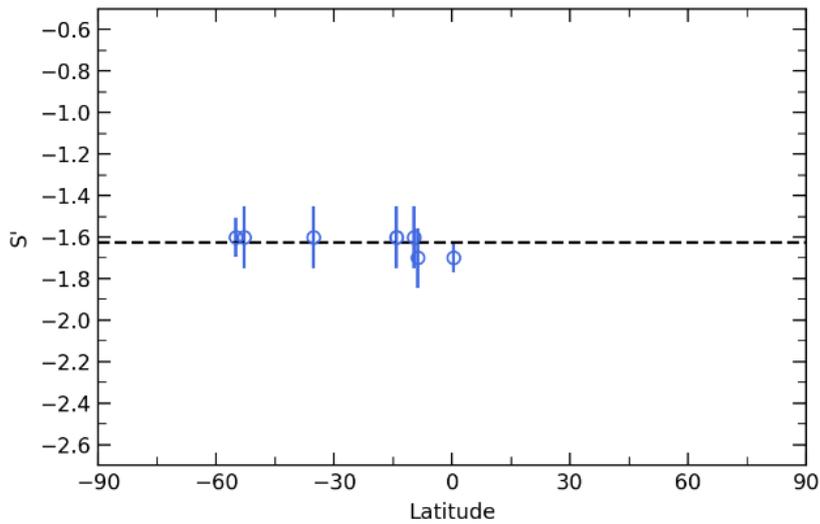

**Figure 6.** Spectral gradient $S'$ according to the latitude of Phaeton. The blue dots indicate $S'$ determined from the observation data, and the dashed lines represent the mean values of $S'$.

In order to explain why latitude-dependent color variation does not appear, we assumed that the surface of Phaethon were experienced decomposition from hydrated C-type to F/B-type by thermal metamorphism as it passes close to the Sun. Hence, we re-calculated the solar-radiation heating effect, which was previously calculated by Ohtsuka et al. (2009), using the solution and shape model of Kim et al. (2018). For the ephemeris calculation the JPL Horizons service was used, and the asteroid sub-solar point temperature was determined based on the energy balance equation on an asteroid's surface,

$$T_{SS} = \left[\frac{S_\odot}{r^2} \frac{(1-A)}{\eta \varepsilon \sigma}\right]^{1/4} \quad (2)$$

where A is the bond albedo, $A \approx (0.290 + 0.684G) p_V$, and we used $G = 0.15 \pm 0.03$ and $p_V = 0.12 \pm 0.01$ (Hanus et al. 2016). $S_\odot$ denotes the solar constant, solar flux at 1 $AU$, for which $\sim 1366\ Wm^{-2}$ was used. And $\varepsilon$ is the infrared emissivity and was assumed to be 0.9. In addition, $\eta$ is the beaming parameter at the solar phase angle of $\alpha = 0°$, for which the mean value for a B-type asteroid of $1.0 \pm 0.1$ was used (Alí-Lagoa et al. 2013).

The local surface temperature was calculated using the following equation:

$$T_{loc} = T_{SS} \cos \theta_i^{1/4} \quad (3)$$

where $\theta_i$ is the angular distance between the sub-solar point and the local point. The temperature of the night side was assumed to be 0 K.

To explain the surface properties of Phaethon in term of local surface temperature, we used the thermal metamorphism condition of CI/CM chondrites that was proposed by Nakamura (2005). He distinguished that the thermal metamorphism conditions of CI/CM chondrites occur in four stages according to the surface temperature.

- Stage I (heating < 250 °C; 523 K): This is the lowest heating stage, which is a layered sensitive matrix, and the hydrous minerals of saponite, serpentine and tochilinite exist with no considerable heating effect.

- Stage II (approximately 250–500 °C; 523-773 K): The serpentine is amorphous, but the olivine is recrystallized. No tochilinite exists, and newly created troilite with low crystallinity exists.

- Stage III (approximately 500–750 °C; 773-1023 K): This exists between low crystallinity, fine-grained olivine amorphous matrix phases, and newly created troilite with a low crystallinity exists.

- Stage IV (approximately > 750 °C; 1023-1273 K): The matrix olivine is perfectly anhydrous. The matrix olivine has become well-crystalline, and the newly created troilite is intergrown.

We examined the newly calculated local surface temperatures by dividing them into six regions based on the latitude. The range of each region is shown in Table 4. Fig. 7 shows the surface temperature variations for different regions along the orbit. This figure also confirms that Phaethon's temperature reaches the highest at 15 ° N and it tends to decrease as it moves away from the 15 ° N region. Nevertheless, the latitude temperature distribution of Phaethon suggests that it may have almost compositionally homogeneous surface. At the same time, the 15 ° N region might have partially experienced more regional thermal metamorphism. Our speculation seems to be consistent with our observations. The result of Ohtsuka et al. (2009) conflicts with our simulation. In particular, the surface temperature of its south pole does not reach Stage III and IV in their FRM model. This difference is caused by the different pole orientation they used. The pole orientation is important parameter to determine the incident flux of solar radiation on surface which heavily depends on latitude. Their calculation was based on pole orientation solved by Krugly et al. (2002). However, the pole orientation of Krugly et al. (2002) shows disagreements with subsequent studies (see, e.g. Table 3 of Kim et al. 2018), and our test was calculated employing Kim et al. (2018)'s pole solution. Hence, the uniform color of Phaethon in latitude determined from available spectral data is consistent with the result of our analysis of solar-radiation heating effect. However, spectral data currently available for Phaethon does not fully cover its surface in latitude, and especially miss the northern hemisphere. Therefore, follow-up observations for other apparitions with a different viewing geometry in which the other hemisphere can be observed is required to confirm the homogeneity of the surface.

**Table 4.** Latitude region ranges

| | Latitude Range |
|---|---|
| NP | 60° ~90° |
| 45°N | 30° ~ 60° |
| 15°N | 0° ~ 30° |
| 15°S | -30° ~ 0° |
| 45°S | -60° ~ -30° |
| SP | -90° ~ -60° |

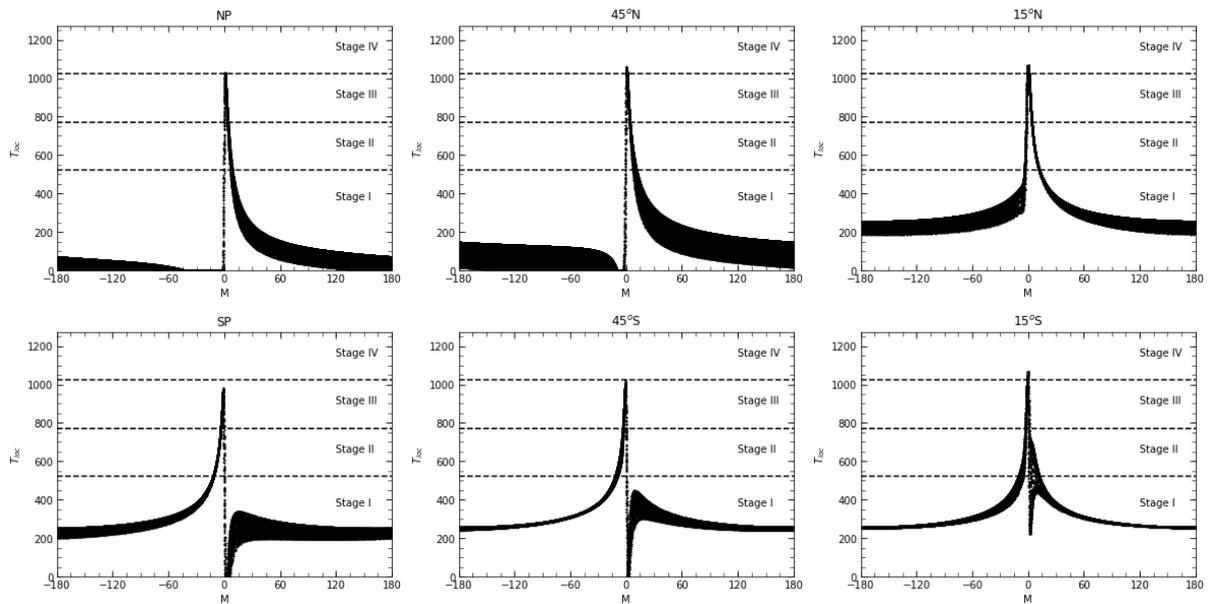

**Figure 7.** The maximum surface temperature, $T_{loc}$, according to the mean anomaly, M, of each region. The dashed lines show the boundary points of each stage as presented in Nakamura (2005).

## 4. Summary and Conclusions

We conducted time-series observations using 4-band photometry and long-slit spectroscopy to verify the taxonomy and homogeneity of Phaethon's surface. The results showed that it is a B-type asteroid, which is consistent with the results of previous studies, and that its southern hemisphere has a homogeneous surface. We also collected archival data and compared them with our dataset which revealed that that there was no surface inhomogeneity in latitude direction. We re-calculated the solar-radiation heating effect for Phaethon using the solution of Kim et al. (2018). We calculated its surface temperature using the energy balance equation on an asteroid's surface, and the result confirmed that the solar-radiation heating effect on Phaethon's mineralogy appears similar at every latitude, implying a latitudinally uniform surface for this asteroid. However, our observation data are still insufficient to cover the entire surface. Therefore, additional time-series multi-band photometry and spectral observations are required to completely understand the surface nature of Phaethon.

**Acknowledgements**

This research is supported by the Korea Astronomy and Space Science Institute (KASI). Part of the data utilized in this publication was obtained and made available by the MIT-UH-IRTF Joint Campaign for NEO Reconnaissance. The IRTF is operated by the University of Hawaii under Cooperative Agreement no. NCC 5-538 with the National Aeronautics and Space Administration, Office of Space Science, Planetary Astronomy Program. The MIT component of this work is supported by NASA grant 09-NEOO009-0001 and by the National Science Foundation under Grants Nos. 0506716 and 0907766.

**Appendix A. Supplementary data**

Supplementary data related to this article can be found at

https://data.mendeley.com/datasets/k3b7s7d6dt/draft?a=f351eeee-7923-492c-a507-7004da1ca5f9.